\documentclass[12pt]{article}
\usepackage{epsf}
\def\beq{\begin{equation}}
\def\eeq{\end{equation}}
\def\ap#1#2#3 {Ann. Phys. (NY) {\bf#1} (19#2) #3}
\def\err#1#2#3 {{\it Erratum} {\bf#1} (19#2) #3}
\def\ib#1#2#3 {{\it ibid.} {\bf#1} (19#2) #3}
\def\ijmp#1#2#3 {Int. J. Mod. Phys. {\bf#1} (19#2) #3}
\def\jetp#1#2#3 {JETP Lett. {\bf#1} (19#2) #3}
\def\mpl#1#2#3 {Mod. Phys. Lett. {\bf#1} (19#2) #3}
\def\np#1#2#3 {Nucl. Phys. {\bf#1} (19#2) #3}
\def\pl#1#2#3 {Phys. Lett. {\bf#1} (19#2) #3}
\def\prep#1#2#3 {Phys. Rep. {\bf#1} (19#2) #3}
\def\prev#1#2#3 {Phys. Rev. {\bf#1} (19#2) #3}
\def\prl#1#2#3 {Phys. Rev. Lett. {\bf#1} (19#2) #3}
\def\sjnp#1#2#3 {Sov. J. Nucl. Phys. {\bf#1} (19#2) #3}
\def\spj#1#2#3 {Sov. Phys. JETP {\bf#1} (19#2) #3}
\def\spu#1#2#3 {Sov. Phys. Usp. {\bf#1} (19#2) #3}
\def\zp#1#2#3 {Zeit. Phys. {\bf#1} (19#2) #3}
\def\a{\alpha}

\begin{document}
\begin{titlepage}
\begin{center}
{\Large \bf William I. Fine Theoretical Physics Institute \\
University of Minnesota \\}  \end{center}
\vspace{0.2in}
\begin{flushright}
FTPI-MINN-04/10 \\
UMN-TH-2238-04 \\
February 2004 \\
\end{flushright}
\vspace{0.3in}
\begin{center}
{\Large \bf  Relative yield of charged and neutral heavy meson pairs in $e^+
e^-$ annihilation near threshold
\\}
\vspace{0.2in}
{\bf M.B. Voloshin  \\ }
William I. Fine Theoretical Physics Institute, University of
Minnesota,\\ Minneapolis, MN 55455 \\
and \\
Institute of Theoretical and Experimental Physics, Moscow, 117259
\\[0.2in]
\end{center}

\begin{abstract}
The subject of the charged-to-neutral yield ratio for $B {\bar B}$ and $D {\bar
D}$ pairs near their respective thresholds in $e^+e^-$ annihilation is
revisited. As previously argued for the $B$ mesons, this ratio should exhibit a
substantial variation across the $\Upsilon(4S)$ resonance due to interference of
the resonance scattering phase with the Coulomb interaction between the charged
mesons. A simple alternative derivation of the expression describing this effect
is presented here, and the analysis is extended to include the $D$ meson
production in the region of the $\psi(3770)$ resonance. The available data on
Kaon production at the $\phi(1020)$ resonance are also discussed in connection
with the expected variation of the charged-to-neutral yield ratio.
\end{abstract}

\end{titlepage}

\section{Introduction}
The near-threshold resonances $\phi(1020)$, $\psi(3770)$, and $\Upsilon(4S)$,
decaying respectively to pairs of pseudoscalar mesons $K {\bar K}$, $D {\bar
D}$, and $B {\bar B}$ are well known `factories' for production of these mesons
in $e^+ e^-$ annihilation. In a number of experimental studies it is important
to know the relative yield of the pairs of charged and neutral mesons in the
corresponding resonance region, i.e. the ratio
\beq
R^{c/n}={\sigma(e^+e^- \to P^+P^-) \over \sigma(e^+e^- \to P^0
{\bar P}^0)}~,
\label{rcn}
\eeq
where $P$ stands for the pseudoscalar meson, i.e. $K$, $D$, or $B$. The
knowledge of this ratio is of particular importance for the studies of the $B$
mesons at the $B$-factories, and dedicated measurements have been done at the
$\Upsilon(4S)$ resonance by CLEO\cite{cleo1,cleo2}, $BABAR$\cite{babar1,babar2}
and Belle\cite{belle}. The central values of the ratio $R^{c/n}$ found in these
measurements typically range from 1.01 to 1.10 with the latest
result\cite{babar2} being $1.006 \pm 0.036(stat) \pm 0.031(sys)$. The same ratio
for the production of $D$ meson pairs is likely to be studied in detail at the
$\psi(3770)$ resonance by the imminent CLEO-c experiment\cite{cleoc}. For the
Kaon pair production the most detailed available scan of production of charged
and neutral mesons in the $\phi$ resonance region has been done by the SND
collaboration\cite{snd}.

The theoretical treatment of the ratio $R^{c/n}$ at these three thresholds near
the corresponding resonances bears considerable similarity, although specific
factors contributing to the ratio in each case are slightly different. In a way,
the simplest case is presented by the threshold $B$ pair production, where the
deviation of $R^{c/n}$ from one is essentially entirely determined by the
Coulomb interaction between the charged $B$ mesons, since the mass difference
between the $B$ mesons is very small: $m_{B^0}-m_{B^+}=0.33 \pm 0.28 \,
MeV$\cite{pdg}. For the $D {\bar D}$ pair production at the $\psi(3770)$ peak
the difference in the $P$-wave  kinematical factor $p^3$ due to the substantial
mass difference between the charged and neutral $D$ mesons is the main source of
deviation of $R^{c/n}$ from one, and the Coulomb interaction effect is somewhat
smaller, but is still of a measurable $O(10\%)$ magnitude. Finally, in the
$\phi(1020)$ resonance region in addition to the mass difference and the Coulomb
interaction effects, the amplitudes of production of $K^+K^-$ and $K_LK_S$ also
differ due to a non-resonant isovector, $I=1$, contribution coming from the
electromagnetic current of the $u$ and $d$ quarks. Another technical difference
between the heavy mesons and the Kaons is that the former can safely be
considered nonrelativistic at energies in the region of the corresponding
resonance (the velocities of the mesons at the corresponding resonances are
$v(B)/c \approx 0.06$ and $v(D^+)/c \approx 0.13$), while the velocity of the
charged Kaons at the $\phi$ resonance is $v(K^+)/c \approx 0.25$, and the
$O(v^2/c^2)$ relativistic effects can show up at some level of intended accuracy
in $R^{c/n}$.

The kinematical effect of the mass difference, and of the non-resonant isovector
amplitude for the case of Kaon production, lead to factors in the $R^{c/n}$,
which are rather smoothly varying within the width of the corresponding
resonance. The behavior of the Coulomb interaction effect is however quite
different. Namely, as recently pointed out\cite{mv}, with a proper treatment of
the strong resonant interaction between the mesons, there is a phase
interference between the Coulomb interaction and the resonance (Breit-Wigner)
scattering phase. Since the latter phase changes by $\pi$ across the resonance,
the effect of the Coulomb interaction in $R^{c/n}$ should exhibit a substantial
variation with energy within the width of the resonance. Thus in particular a
comparison between different measurements of the charged-to-neutral ratio at the
$\Upsilon(4S)$ peak can only be meaningful with proper account for the
differences in the specific experimental setups such as the beam energy spread,
stability, calibration, etc. Needless to mention that the best approach would be
a dedicated scan of the $\Upsilon(4S)$ peak. The details of this variation
depend\cite{mv} on the specifics of the transition from the strong interaction
region at short distances to the long-range Coulomb interaction and also on such
details as the non-resonant scattering phase for the mesons. A dedicated scan of
the charged-to-neutral ratio thus could possibly provide an information on
rather fine properties of strong interaction between the heavy mesons, which are
not likely to be accessible by other means. It may well be that a more
experimentally feasible object for such study is provided by the $\psi(3770)$
resonance, and possibly can be explored in detail in the forthcoming CLEO-c
experiment. The main motivation of the present paper is to extend the analysis
of Ref.\cite{mv} to the case of the $D {\bar D}$ production in the region of
this resonance, and to estimate the magnitude of the expected effects, in the
extent that it can be done within the present (rather approximate at best)
understanding of the details of the strong interaction between heavy mesons.

The typical amplitude of the varying part of the ratio $R^{c/n}$ due to the
Coulomb interaction is set by the Coulomb parameter $\alpha/v$ for the charged
mesons, which tells us that for the $D$ mesons at $\psi(3770)$ the amplitude of
the variation should be approximately twice smaller than for the $B$ mesons at
$\Upsilon(4S)$ (and another factor of two smaller for the Kaons at $\phi$). In
either case the Coulomb parameter is sufficiently small to justify limiting the
consideration to only the linear term in the Coulomb interaction treated as
perturbation. A simple derivation of the linear in the Coulomb interaction
correction to production of a charged meson pair with the exact treatment of the
strong interaction scattering is presented in Section 2. This derivation is
slightly different from that given in Ref.\cite{mv}, and hopefully, more
transparently leads to the expression for the $O(\alpha/v)$ term found there.
In Section 3 are presented specific estimates of the behavior of the ratio
$R^{c/n}$ across the $\psi(3770)$ resonance, which take into account both the
mass difference between the $D$ mesons and the effect of the Coulomb
interaction.

It can be noticed that the very fact of the existence of the variation of
$R^{c/n}$ within the resonance width is essentially model independent and is a
direct consequence of a somewhat standard application of the quantum mechanical
scattering theory. Nevertheless, there has been some scepticism expressed
regarding the existence of the predicted\cite{mv} rapid variation of $R^{c/n}$
at the resonance, most notably in the introductory part of the recent
experimental paper\cite{babar2}. Although the reported in Ref.\cite{babar2}
measurement has been done at just one point in the energy at the peak of the
$\Upsilon(4S)$ resonance, the paper implicitly puts the prediction of the
variation in doubt by arguing that ``such rapid variation in the
charged-to-neutral ratio has not been observed in scans across the $\phi(1020)$
resonance" (with a reference to the scans by the SND collaboration\cite{snd}).
Thus it appears to be useful to point out the difference between the predictions
for the $B$ production at $\Upsilon(4S)$ and the Kaon production at $\phi(1020)$
following from the discussed approach. The main difference is that the amplitude
of the variation at the $\phi$ is expected to be suppressed by a factor of
approximately 0.25 due to about four times larger velocity of the Kaons at the
relevant energy, and additional differences in the details arise through
different parameters (e.g. widths) of the $\phi$ and $\Upsilon(4S)$ resonances.
A comparison of a `representative' theoretical curve with the behavior of
$R^{c/n}$ across the $\phi$ resonance extracted from the scan data\cite{snd} is
presented in Section 4. Given the existing theoretical and experimental
uncertainties, a meaningful comparison can only be done at a mostly qualitative
level, and it is left entirely up to the reader to assess from the presented
comparison whether the scan data indeed exclude a variation with expected
amplitude of the charged-to-neutral ratio across the $\phi$ resonance.

\section{Coulomb interaction correction to production of charged mesons}

If the mesons were point particles devoid of strong interaction and produced by
a point source, the Coulomb interaction between charged mesons would enhance
their production cross section by the textbook factor
\beq
F_c=1+{\pi \, \alpha \over 2 v} + O\left ( \alpha^2/v^2 \right )~,
\label{fc}
\eeq
where $v$ is the velocity of each of the mesons in the center of mass frame,
which was the early prediction\cite{am} for the ratio $R^{c/n}$ at the
$\Upsilon(4S)$ resonance. It was however then realized that the effects of the
meson electromagnetic form factor\cite{lepage} and of the form factor in the
decay vertex $\Upsilon(4S) \to B {\bar B}$\cite{be,kmm} tend to reduce the
enhancement. In all these approaches the Coulomb rescattering of the produced
mesons is considered, assuming that the propagation of the meson pair between
the production vertex and the rescattering as well as after the Coulomb
rescattering is described by a free propagator, and thus ignoring the fact that
the strong interaction introduces the scattering phase factors $\exp(i \delta)$
in these propagators. This would be a reasonable approximation if the strong
phase $\delta$ were small. In the region of a strong resonance however this
assumption is definitely invalid, since the phase changes by $\pi$ within the
width of the resonance. The calculation described in Ref.\cite{mv} allows to
completely take into account the strong scattering phase as well as the Coulomb
interaction between the mesons. In the first order in the Coulomb rescattering
the expression\cite{mv} for the Coulomb correction factor has a simple form
\beq
F_c=1+{ 1 \over v} \, {\rm Im}\left [ e^{2i \delta} \, \int_a^\infty e^{2ipr} \,
\left ( 1+ {i \over p r} \right
)^2 \, V(r) \, dr \right ]~,
\label{fcm}
\eeq
where $p$ is the meson momentum and $V(r)$ is the potential for the rescattering
interaction. Clearly, up to a possible form factor, $V(r)=-\alpha/r$ for the
Coulomb interaction. The short-distance cutoff parameter $a$ in eq.(\ref{fcm})
accounts for the fact that in the region of a strong interaction at short
distances the mesons spatially overlap and in fact the system is a mixture of
heavy and light quarks and gluons rather than a state of two individual mesons.
Therefore at such distances there is no separation between the states with
charged and neutral mesons and thus any difference in their Coulomb interaction
disappears. If one introduces form factors in the Coulomb interaction in order
to account for the structure of the mesons, the interaction potential $V(r)$
would automatically vanish at short distances and the ultraviolet cutoff in the
integral in eq.(\ref{fcm}) would be provided by the form factors. If as in the
earlier analyses, one sets the strong scattering phase $\delta$ to zero, the
expression in eq.(\ref{fcm}) is finite even if an unmodified Coulomb potential
is taken down to $a=0$, in which limit it reproduces eq.(\ref{fc}). However at
any finite $\delta$ there is an essential dependence on the ultraviolet cutoff
parameter $a$ which cannot be removed.

In this section we present a slightly different from that in Ref.\cite{mv}
derivation of eq.(\ref{fcm}) by considering the following problem. Two mesons,
each with mass $m$ interact strongly at distances $r < a$. The strong
interaction in the $P$ wave at the total kinetic energy $E$ produces the
scattering phase $\delta$. At longer distances, $r > a$ the mesons interact via
the potential $V(r)$. The meson pair is produced by a source localized at
distances shorter than $a$ (this obviously corresponds to the production in the
$e^+e^-$ annihilation). The problem is to find the correction of the first order
in $V$ to the production rate.

In order to solve the formulated problem we consider the radial part of the wave
function of a stationary $P$-wave scattering state, written in the form
$R(r)=\chi(r)/r$. According to the general scattering theory\cite{ll} the
asymptotic form of $\chi(r)$ at $r \to \infty$ in the absence of the long-range
potential $V(r)$ is
\beq
\chi(r) = 2 \cos(pr + \delta)= e^{i\delta} \, e ^{i pr} + e^{-i\delta} \, e ^{-i
pr}~.
\label{chias}
\eeq
Furthermore, in the absence of the potential $V$ the function $\chi(r)$
satisfies the Schr\"odinger equation for free motion in the $P$ wave:
\beq
\chi''(r)+\left (p^2 - {2 \over r^2} \right ) \, \chi(r)
=0
\label{sch0}
\eeq
at all distances $r$ outside the region of strong interaction, i.e. at $r > a$.
Thus the solution valid at all such distances, and having the asymptotic form
(\ref{chias}) can be written explicitly:
\beq
\chi(r)=e^{i\delta} \, f( pr) + e^{-i\delta} \, f^*( pr)~,
\label{chi0}
\eeq
in terms of the function
\beq
f(pr)=\left ( 1+ {i \over p r} \right ) e^{ipr}~,
\label{fr}
\eeq
and of its complex conjugate $f^*$. At distances $r < a$ the dynamics is unknown
to the extent that continuing description of the system in terms of the meson
pair wave function in the region of strong interaction does not make much sense,
since most likely such description should involve entirely different degrees of
freedom. Nevertheless whatever complicated the `inner' dynamics of the system
may be, the wave function in eq.(\ref{sch0}) at $r=a$ provides the boundary
condition for the `inner' problem. In particular, it determines the
normalization of the state for the `inner' problem. Thus the rate of production
of the system, and consequently of the meson pair, by a source localized at $r
\ll a$ is proportional to $|\chi(a)|^2$.

When the long-range potential $V(r)$ is turned on and considered as a
perturbation the wave function at $r > a$ changes to $\chi(r)+\delta\chi(r)$.
Once the corrected solution is normalized at $r \to \infty$ in the standard way,
i.e. as the asymptotic wave in eq.(\ref{chias}), the normalization at $r=a$
generally changes and hence also changes the production rate. In the first order
in $V$ the correction factor for the production rate is obviously given by
\beq
F_c=1+2 {\delta \chi(a) \over \chi(a)}~.
\label{fcorr}
\eeq

The perturbation $\delta \chi$ of the scattering state wave function can be
found in the standard way by writing it as a sum of outgoing and incoming waves:
$\delta \chi(r) = \delta \chi_+( r) +  \delta \chi_-( r)~,$  similarly to
eq.(\ref{chi0}), and where $\delta \chi_+(r)$ contains at $r \to \infty$ only an
outgoing wave ($\exp(i p r)$) while $\delta \chi_-=\delta \chi_+^*$ contains at
asymptotic $r$ only an incoming wave. The function $\delta \chi_+(r)$ is a
perturbation of the outgoing wave part of the function $\chi(r)$ in
eq.(\ref{chi0}) and is thus determined by the equation
\beq
\delta \chi_+''(r)+\left (p^2 - {2 \over r^2} \right ) \, \delta \chi_+(r)=m \,
V(r) \, e^{i\delta} \, f( pr)~.
\label{schpert}
\eeq

The solution to this equation is found by the standard method using the Green's
function $G_+(r,r')$ satisfying the equation
\beq
\left ( {\partial^2 \over \partial r^2} + p^2 - {2 \over r^2} \right ) \,
G_+(r,r') = \delta(r-r')~,
\label{gfe}
\eeq
and the condition that $G_+(r,r')$ contains only an outgoing wave when either of
its arguments goes to infinity. The Green's function is constructed from two
solutions of the homogeneous equation, i.e. from the functions $f(pr)$ and
$f^*(pr)$, as
\beq
G_+(r,r')={1 \over 2 \, i \, p} \left[ f(pr) \, f^*(pr') \, \theta(r-r') +
f(pr') \, f^*(pr) \, \theta(r'-r) \right ]~,
\label{gsol}
\eeq
where $\theta$ is the standard unit step function. The solution to the equation
(\ref{schpert}) is then found as
\beq
\delta \chi_+(r)=m \, e^{i\delta} \,\int_a^\infty G_+(r,r') \, V(r') \, f(pr')
\, dr'~.
\label{chisol}
\eeq
It can be noted that at no point in this consideration a knowledge of the
dynamics at $r < a$ is required. In particular the integral in eq.(\ref{chisol})
runs from the lower limit at $a$, since by our assumptions the perturbation
potential has support only at $r > a$.

It is further important that adding the found perturbation $\delta \chi$ to the
wave function does not change the normalization of the wave function at $r \to
\infty$. Indeed in this limit one has $r > r'$ in the Green's function in
eq.(\ref{chisol}), so that the asymptotic behavior of $\delta \chi_+(r)$ should
be derived from the expression
\beq
\delta \chi_+(r) \left. \right |_{r \to \infty} = -{i \over 2 v} \, e^{i\delta}
\, f(pr) \int_a^\infty V(r') \, |f(p r')|^2 \, dr'~,
\label{chipi}
\eeq
which gives the complex phase of $\delta \chi_+(r)$ at the asymptotic distances
manifestly orthogonal to that of the outgoing-wave part of $\chi(r)$, since the
integral is explicitly real\footnote{A minor technical point is that formally
the integral is logarithmically divergent at $r' \to \infty$ for the Coulomb
potential, which might put into question the applicability of the condition $r >
r'$ in the entire integration region. This is the standard infrared divergence
of the Coulomb scattering phase, and it can be easily dealt with by temporarily
introducing a small photon mass $\lambda$, so that the potential is cut off by
the factor $\exp(-\lambda r)$ and the integral is convergent. In the final
result one can obviously take the limit $\lambda \to 0$.}. In other words, at $r
\to \infty$ the correction changes only the scattering phase by adding the
`Coulomb' phase to $\delta$.

At $r = a$ the perturbation $\delta \chi_+(a)$ is found from the same expression
(\ref{chisol}). In this case one has $r < r'$ in the entire integration region,
so that
\beq
\delta \chi_+(a )  = -{i \over 2 v}\, e^{i\delta} \, f^*(pa) \int_a^\infty V(r')
\, \left [f(p r') \right]^2 \, dr'~.
\label{dchia}
\eeq
Using this expression one readily finds the change in the normalization of the
wave function at $r=a$ due to the potential $V$, and thus arrives at the
expression (\ref{fcm}).

As is clear from the presented derivation, the result in eq.(\ref{fcm}) is valid
at arbitrary $P$-wave scattering phase $\delta$, and also generally does not
assume any condition for the value of the product $p a$. At energy $E$ near a
strong resonance dominating the dynamics of the meson pair, the scattering phase
is determined by the Breit-Wigner formula
\beq
\delta= \delta_{BW}+\delta_1,~~~e^{2i\delta_{BW}}={\Delta - i \gamma \over
\Delta + i \gamma}~,
\label{dbw}
\eeq
where $\Delta=E-E_0$ is the distance in energy to the nominal position $E_0$ of
the resonance, and $\gamma$ is generally a function of energy such that its
value at $E_0$ is related to the nominal width $\Gamma$ of the resonance as
$\gamma(E_0)=\Gamma/2$. Finally, $\delta_1$ in eq.(\ref{dbw}) is the
non-resonant $P$-wave scattering phase.

The properties of the discussed resonances, $\Upsilon(4S)$ and $\psi(3770)$ are
determined by the two meson-pair channels: $P^+ P^-$ and $P^0 {\bar P}^0$. The
analytical properties of the scattering amplitude in the $P$ wave require that
the contribution of each channel to $\gamma$ and $\delta_1$ starts at the
corresponding meson pair threshold as cubic in the corresponding momentum. Thus
at the energy close to both thresholds one can parametrize these quantities as
\beq
\gamma(E)=c_1 \, (p_+^3+p_0^3)~,~~~~ \delta_1(E)=c_2 \, (p_+^3+p_0^3)~,
\label{gd}
\eeq
where $p_+$ and $p_0$ are the momenta of respectively the charged and the
neutral mesons at the energy $E$. Due to the very small mass difference between
the $B$ mesons, the thresholds for the charged and neutral $B$ mesons
practically coincide and so do the momenta $p_+$ and $p_0$. However for the $D
{\bar D}$ pairs the $D^+ D^-$ threshold is by $9.6 \, MeV$ higher than that for
$D^0 {\bar D}^0$, which difference is substantial at the energy of the
$\psi(3770)$ resonance.

It should be noticed that neglecting the higher terms of expansion in the
momenta for the parameters $\gamma$ and $\delta_1$ is justified only inasmuch as
the momenta are small in the scale of the size of the strong interaction region
$a_s$. In the calculation of the Coulomb interaction effect it is essential that
the distance parameter $a$ for the onset of the Coulomb interaction is not
smaller than $a_s$: $a \ge a_s$. Otherwise no restriction on the value of the
product $pa$ is implied. In Ref.\cite{mv} these parameters were reasonably
assumed to be approximately equal. However it should be emphasized that
generally the applicability of the formula (\ref{fcm}) for the Coulomb
correction is separate from the applicability of the first terms of the
threshold expansion in eq.(\ref{gd}), and these two issues can be studied
separately in experiments\footnote{In fact an attempt at detecting higher than
$p^3$ terms in the width parameter for the $\Upsilon(4S)$ resonance has been
done by ARGUS collaboration\cite{argus}. However the deviation from the cubic
behavior turned out to be too small within the experimental accuracy.}.

In conclusion of this section we write the explicit formula for the correction
described by eq.(\ref{fcm}) in the resonance region in terms of real
quantities\cite{mv}:
\beq
F_c=1+{\a \over v} \, \left [ {\Delta^2 - \gamma^2 \over \Delta^2 +
\gamma^2}\, \left ( A \, \cos 2 \delta_1 + B \, \sin 2 \delta_1 \right )
-  {2 \, \gamma \, \Delta \over \Delta^2 + \gamma^2} \, \left ( B\, \cos
2 \delta_1 - A \, \sin 2 \delta_1 \right ) \right ]~.
\label{fcw}
\eeq
The coefficients $A$ and $B$ are found as the imaginary and the real parts of
the integral in eq.(\ref{fcm}) with the Coulomb potential:
\begin{eqnarray}
&&A = -\int_{pa}^\infty \left [  \left (1-{1 \over u^2} \right )\,\sin
2u  + {2 \, \cos 2u \over u} \right ] \, {du \over u}={\pi \over 2}-{\cos 2 pa
\over pa} + {\sin 2 pa \over 2 (pa)^2} - Si(2pa),\nonumber \\
&&B = \int_{pa}^\infty \left [ {2 \,\sin 2 u \over u}- \left (1- {1
\over
u^2} \right ) \, \cos 2 u  \right ] \, {du \over u}={\cos 2 pa \over 2
(pa)^2}+{\sin 2 pa \over pa}-Ci(2pa) ~,
\label{ab}
\end{eqnarray}
where
$$ Si(z)=\int_0^z \sin t \, {dt \over t}~~~~{\rm and}~~~~Ci(z)=-\int_z^\infty
\cos t \, {dt \over t}~.$$

\section{Estimates of the expected variation of $R^{c/n}$ across the
$\psi(3770)$ resonance}

Here we apply the formulas of the previous section, specifically the equations
(\ref{fcw}), (\ref{ab}), and (\ref{gd}), to an estimate of the scale of the
variation in the ratio $R^{c/n}$ for the $D {\bar D}$ pair production at the
$\psi(3770)$ resonance. In doing so it should be clearly understood that at
present one can only guess (in some `reasonable' range) the appropriate values
of the scattering phase $\delta_1$ and the cutoff parameter $a$ for the Coulomb
interaction (or more generally make a guess about an appropriate model for the
onset at short distances of the Coulomb interaction between the charged $D$
mesons).

As different from the case of the $B$ meson production at the $\Upsilon(4S)$, in
the charged-to-neutral ratio for the $D$ meson production at $\psi(3770)$ the
Coulomb effect multiplies the ratio of the $p^3$ factors:
\beq
R^{c/n}=F_c \, \left ( p_+ \over p_0 \right )^3~.
\label{rcnd}
\eeq
Also, according to the Particle Data Tables\cite{pdg}, the width of $\psi(3770)$
is somewhat larger than that of the $\Upsilon(4S)$: $\Gamma(\psi'') = 23.6 \pm
2.7 \, MeV$ as opposed to $14 \pm 5 \, MeV$, which tends to smoothen the
variation of $R^{c/n}$ near the $\psi(3770)$ peak.
\begin{figure}[ht]
\begin{center}
 \leavevmode
    \epsfxsize=8cm
    \epsfbox{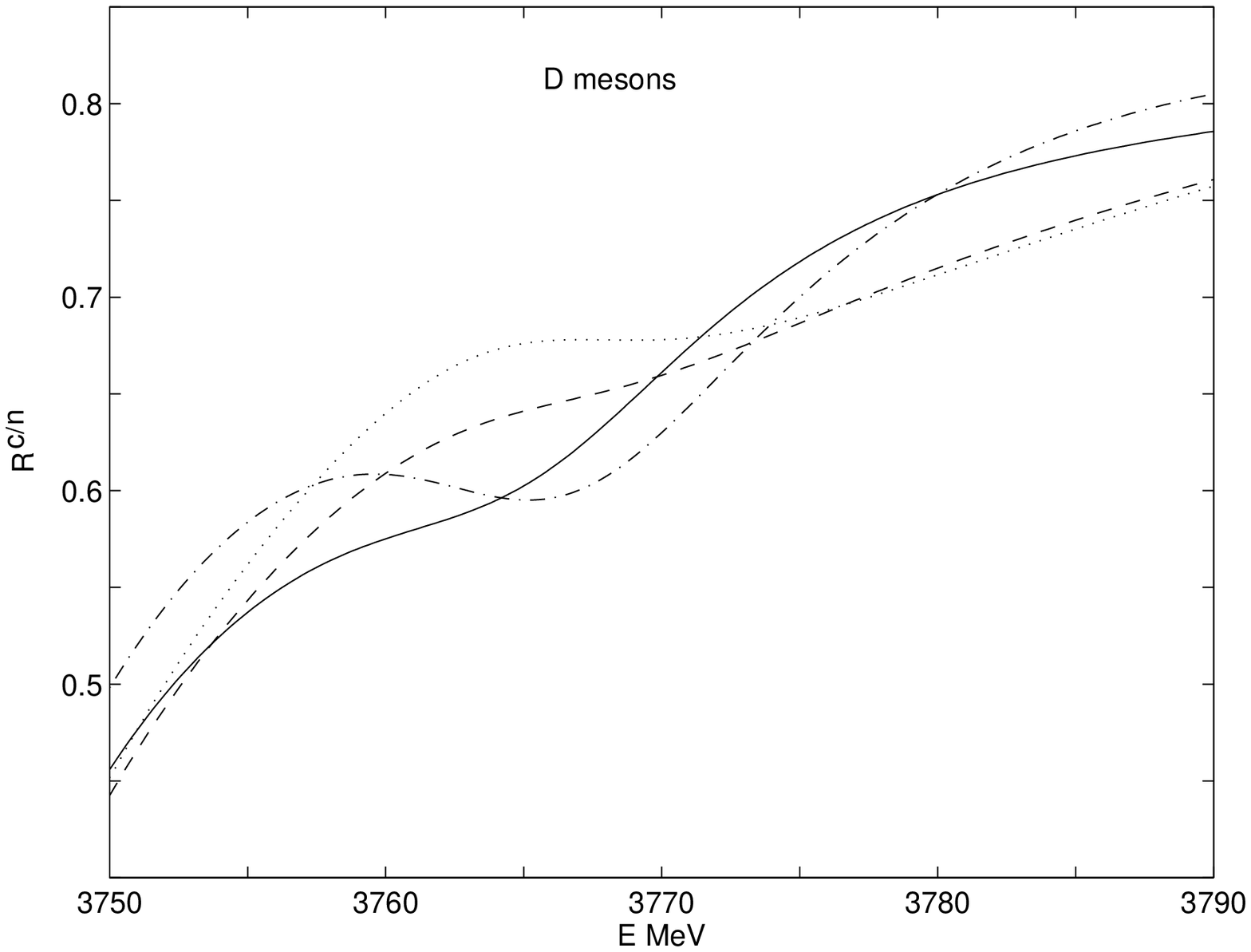} 
    \epsfxsize=8cm
    \epsfbox{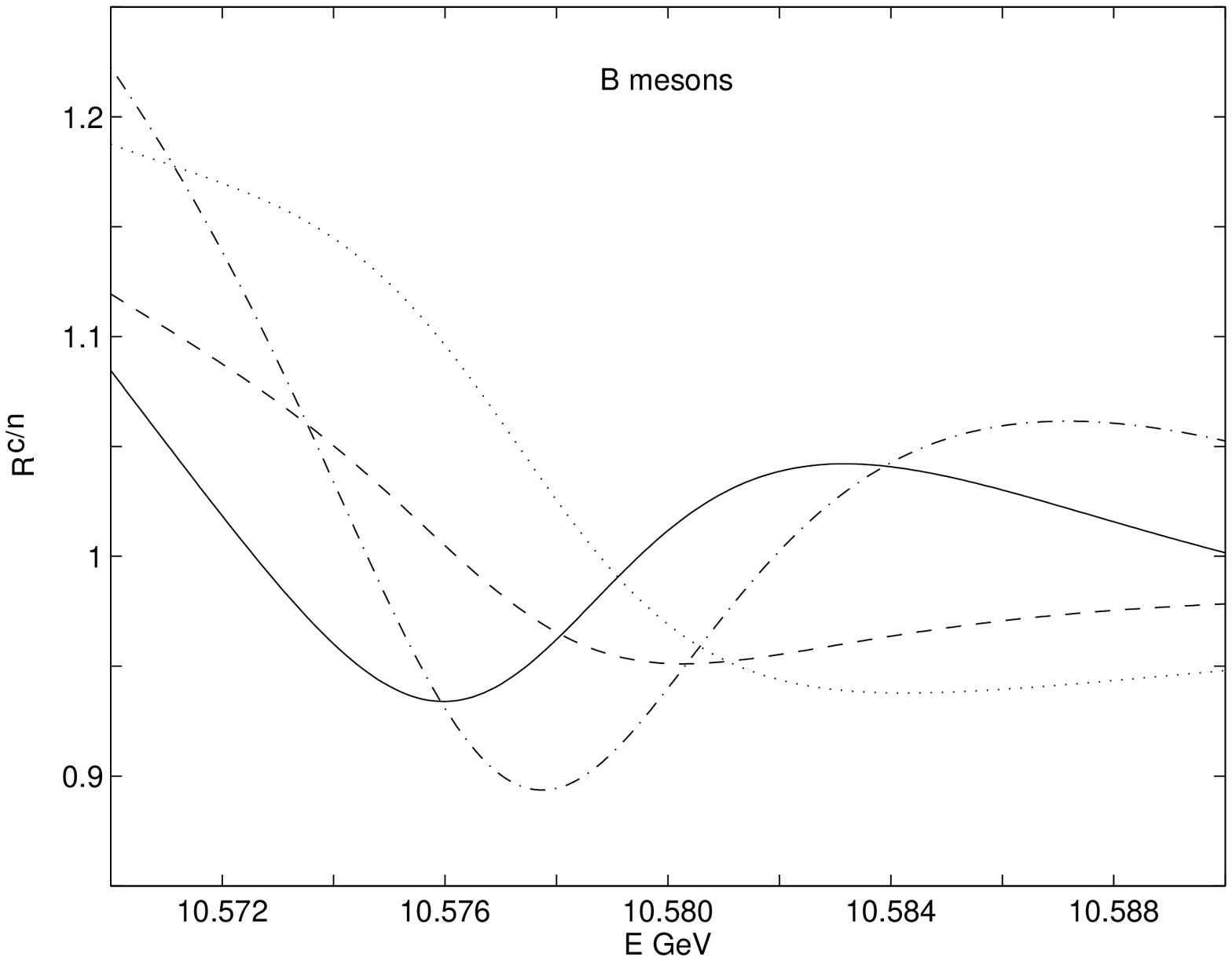}
    \caption{The energy dependence of the ratio $R^{c/n}$ for $D$ and $B$ meson
pair production in the region of the respective resonance: $\psi(3770)$ and
$\Upsilon(4S)$
(the center positions are assumed at $E_0=3.770 \, GeV$ and $E_0=10.580 \, GeV$
respectively ) for some
values of $a$ and $\delta_1(E_0)$: $a^{-1}=200 \, MeV,~\delta_1(E_0)=30^0$
(solid), $a^{-1}=200 \, MeV,~\delta_1(E_0)=-30^0$ (dashed),
$a^{-1}=300 \, MeV,~\delta_1(E_0)=30^0$ (dashdot), and $a^{-1}=300 \,
MeV,~\delta_1(E_0)=-30^0$ (dotted).}
\end{center}
\end{figure}

A sample expected behavior of the ratio $R^{c/n}$ for the $D$ pair production
near the $\psi(3770)$ resonance is shown in Fig.1 for some `representative'
values of the parameters $a$ and $\delta_1(E_0)$. For comparison the same curves
for the $B$ pairs in the vicinity of the $\Upsilon(4S)$ peak are also shown in a
separate plot in Fig.1. It should be noted in connection with this comparison
that generally it is not expected that the phase $\delta_1$ is the same in these
two cases, although the parameter $a$ viewed as characterizing the electric
charge structure in a heavy meson is likely to be quite similar for $D$ and $B$
mesons, if one considers both $c$ and $b$ quarks as asymptotically heavy. With
all the present uncertainty in the knowledge of the strong interaction
parameters, one can see from this comparison that the expected variation of the
ratio $R^{c/n}$ is quite less prominent for $D$ mesons at the $\psi(3770)$ peak
than for $B$ mesons at the $\Upsilon(4S)$. Hopefully, the amplitude of the
variation of $R^{c/n}$ at the $\psi(3770)$ is still sufficient for a study in
the upcoming CLEO-c experiment.

\section{Discussion of the $\phi(1020)$ data and summary}

So far the most detailed scan data for the production cross section of charged
and neutral mesons are available only for the Kaons in the vicinity of the
$\phi(1020)$ resonance\cite{snd}, and it is quite natural to discuss whether any
hint at the discussed variation of $R^{c/n}$ is indicated by those data. A
consideration of this ratio at the $\phi$ peak however encounters certain
peculiarities.
As mentioned above, the Coulomb interaction effect is relatively smaller for
production of $K$ mesons at the $\phi$ resonance, and also the relativistic
effects can play a certain role. The analysis of the ratio $R^{c/n}$ is further
complicated by the fact that the production amplitude also receives an isovector
contribution, which has opposite sign for $K^+K^-$ and $K_LK_S$ and thus changes
the charged-to-neutral ratio. (In the vector dominance model this contribution
is considered as the `tail' of the $\rho$ resonance.) Neglecting a smooth
variation of the non-resonant $I=1$ amplitude and also of the non-resonant part
of the $I=0$ amplitude (e.g. the `tail' of the $\omega$ resonance) one can
approximate the formula for $R^{c/n}$ in this case as
\beq
R^{c/n}=F_c \, \left ( p_+ \over p_0 \right )^3 \left | {1 + (A_0 - A_1) (\Delta
+ i \gamma) \over 1 + (A_0 + A_1) (\Delta + i \gamma)} \right |^2~,
\label{rcnk}
\eeq
where $A_1$ and $A_0$ stand for the appropriately normalized relative
contribution of the non-resonant $I=1$ and $I=0$ amplitudes. The charged $K$
mesons are lighter than the neutral ones, thus the cube of the ratio of the
momenta is larger than one and decreases towards one at energies far above the
threshold. The data\cite{snd} however display a slight general increase of the
ratio with energy, which shows that the effect of $A_1$ is not negligible.
This makes any `absolute' prediction of $R^{c/n}$ in this case rather
troublesome, as well as of its general variation with energy on a scale somewhat
larger than the width of $\phi$. A real fit of the involved parameters,
including the discussed here rapid variation of the Coulomb factor $F_c$
requires knowledge of the raw data and of the correlation in the errors. For
this reason here in Fig.2 is shown a comparison of the data sets listed in
Ref.\cite{snd} with a `representative' theoretical curve.
\begin{figure}[ht]
  \begin{center}
    \leavevmode
    \epsfxsize=10cm
    \epsfbox{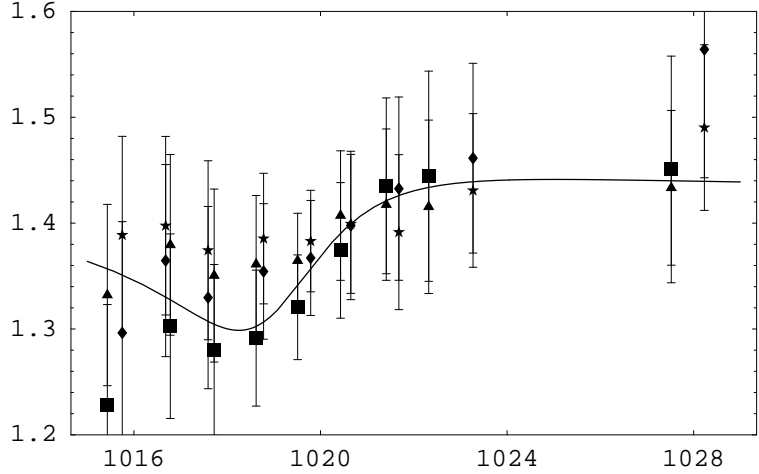}
    \caption{The SND data\cite{snd} at the $\phi(1020)$ resonance represented as
the charged-to-neutral yield ratio for the two reported in \cite{snd} scans
(PHI9801 and PHI9802) and for two methods of identifying the $K_S$ mesons by
their decay into neutral or into charged pions: PHI9801 neutral - stars, PHI9801
charged - diamonds, PHI9802 neutral - triangles, PHI9802 charged - squares. The
solid curve is explained in the text.}
  \end{center}
\end{figure}

More specifically, in the paper \cite{snd} are listed (in the Table IX) the scan
data for the cross section for production of separately the $K^+K^-$ pairs and
the $K_LK_S$ pairs acquired in two different scans (called PHI9801 and PHI9802).
Furthermore the detection of the $K_S$ mesons was done by two separate methods,
i.e. by detecting their decay into charged pions, and into neutral pions.
Accordingly the paper \cite{snd} lists two separate sets of data for the
$K_LK_S$ production measured by each of these two methods.  The systematic error
for the $K^+K^-$ cross section is listed as 7.1\%, and for the neutral Kaons as
4\% and 4.2\% for the two identification techniques used. The statistical errors
are listed for each individual entry for the cross section.  The `data points'
in the plot of Fig.2 are the charged-to-neutral ratia calculated from the data
listed in Ref.\cite{snd} with the errors corresponding to the listed statistical
errors only. The `representative' theoretical curve corresponds to $1.38 F_c$.
In other words the absolute normalization is set rather arbitrarily,
given the described theoretical uncertainty in calculating the absolute
normalization $R^{c/n}$ at the $\phi$ resonance and given the systematic
uncertainty in the data. Also any overall variation of the kinematical and the
amplitude factors in the shown energy interval is neglected. Changing the
normalization of the curve and also including the overall energy dependence
would result in a vertical shift and a slight tilt of the curve. The behavior of
the Coulomb correction factor in the curve shown in Fig.2 is calculated assuming
$E_0=1019.5 \, MeV$, $\Gamma=4.26 \, MeV$, $a^{-1} = 200 \, MeV$, and
$\delta(E_0)=40^0$. The only purpose of the `theoretical' curve in Fig.2 is to
illustrate the approximate magnitude of the expected effect of the variation of
$R^{c/n}$ at the $\phi(1020)$ resonance, and by no means it should be considered
as a detailed prediction.

Due to large experimental and theoretical uncertainties as explained in the text
and can be seen from Fig.2, it is not entirely clear whether the available
data\cite{snd} exclude or rather suggest a variation with the expected magnitude
of the ratio $R^{c/n}$ within the width of the $\phi$ resonance. Perhaps a
detailed global fit of the raw data could provide a statistically significant
evaluation of the amplitude of such variation.

In summary. The strong interaction phase significantly modifies the behavior of
the cross section for production of pairs of charged pseudoscalar mesons near
threshold in $e^+e^-$ annihilation. A simple calculation of this effect in the
first order in the Coulomb interaction is presented in Section 2. The existence
of a strong near-threshold resonance gives rise to a rapid variation of the
charged-to-neutral yield ratio due to the interference of the resonance
scattering phase with the Coulomb phase. The specific behavior of this variation
is sensitive to details of the structure of the mesons and of their strong
interaction. Thus measuring the discussed effect for $B$ mesons at the
$\Upsilon(4S)$ resonance and for $D$ mesons at the $\psi(3770)$ resonance might
provide an information on these details, which are not accessible by other
means. The estimated effect for $D$ mesons in the region of the $\psi(3770)$ is
less prominent than for $B$ mesons at the $\Upsilon(4S)$ but possibly is still
measurable. The expected effect of the rapid variation of the charged-to-neutral
ratio for the Kaon production at the $\phi(1020)$ resonance is the smallest,
and, conservatively, the available data appear to be not conclusive enough to
confirm or exclude such variation with a statistical significance.

\section*{Acknowledgments}
I thank J. Rosner for stimulating questions regarding the estimates of the
behavior of $R^{c/n}$ at the $\psi(3770)$ resonance and A. Vainshtein for a
useful discussion of the Kaon production in the $\phi(1020)$ region.
This work is supported in part by the DOE grant DE-FG02-94ER40823.

\end{document}